\documentclass[a4paper]{jpconf}
\usepackage{epsfig}

\begin{document}
\title{An effective model of QCD thermodynamics\footnote{Talk presented at the Symposium: On Discovery Physics at the LHC -- Kruger 2012, RPA, December~3~-~7,~2012}}

\author{L.~Turko$^1$, D. Blaschke$^{1,2}$, D.~Prorok$^1$ and J.~Berdermann$^3$}
\address{$^1$Instytut Fizyki Teoretycznej, Uniwersytet Wroc{\l}awski, Poland,\\
$^2$Bogoliubov Laboratory for Theoretical Physics, JINR Dubna, Russia,\\
$^3$DESY Zeuthen, Germany}
\ead{turko@ift.uni.wroc.pl, blaschke@ift.uni.wroc.pl, prorok@ift.uni.wroc.pl, jens.berdermann@gmail.com}


\begin{abstract}
A combined effective model reproducing the equation of state of hadronic matter
as obtained in recent lattice QCD simulations is presented.
The model reproduces basic physical characteristics encountered in dense
hadronic matter in the quark-gluon plasma (QGP) phase and the lower
temperature hadron resonance gas phase.
The hadronic phase is described by means of an extended Mott-Hagedorn
resonance gas while the QGP phase is described by the extended PNJL model.
The dissociation of hadrons is obtained by including the state dependent
hadron resonance width.
\end{abstract}

\section {Introduction}

Simulations of lattice QCD (LQCD) are in practice the only reliable approach
to QCD thermodynamics which covers the  broad region of strongly interacting
matter properties from the hadron gas at low temperatures to a deconfined
quark gluon plasma phase at high temperatures.
Recently,  finite temperature LQCD simulations have overcome the difficulties
of reaching the low physical light quark masses and approaching the continuum
limit which  makes this  theoretical  laboratory now  a  benchmark for modeling
QCD under extreme conditions.

We are going to present a combined effective model reproducing the equation of
state of hadronic matter as obtained in recent lattice QCD simulations
\cite{Borsanyi:2010cj,Bazavov:2009zn}.
The model should reproduce basic physical characteristics of processes
encountered in dense hadronic matter, from the hot QCD phase through the
critical temperature region till the lower temperature hadron resonance gas
phase.
In medium properties of hadrons are different from those in the vacuum.
The very notion of the mass shell should be modified, as was postulated quite
long ago \cite{KRT_93}.
The interaction becomes effectively nonlocal due to the Mott effect and
hadrons eventually gradually dissolve into quarks and gluons in the
high temperature phase.
Then, with the increasing temperature,  quark masses are less and less
important although the massless Stefan-Boltzmann limit would be eventually
reached only at extremely high temperature.

It has been shown that the equation of state derived from that time QCD
lattice calculation \cite{Karsch:2001cy} can be reproduced by a simple hadron
gas resonance model.
The rapid rise of the number of degrees of freedom in lattice QCD data
around the critical temperature $T_c \sim 150 - 170$ MeV, can be explained
quantitatively by a resonance gas below the critical temperature $T_c$
\cite{Karsch:2003vd, Ratti:2010kj}.

For higher temperatures the model is modified by introducing finite widths of
heavy hadrons \cite{Blaschke:2003ut, Blaschke:2005za} with a heuristic ansatz
for the spectral function which reflects medium modifications of hadrons.
This fits nicely the lattice data, also above $T_c$ as is shown in
Fig.~\ref{Fig.1}.

This Mott-Hagedorn type model  \cite{Turko:2011gw} has been constructed to fit
nicely the lattice data, also above $T_c$ where it does so because it leaves
light hadrons below a mass threshold of $m_0=1$ GeV unaffected.
The description of the lattice data at high temperatures is accidental because
the effective number of those degrees of freedom approximately coincides with
that of  quarks and gluons.
The QGP presence in this region is formally simulated here by the smart choice
of the mass cut-off parameter such that cut-off defined stable light hadrons
provide the same number of degrees of freedom as partonic components of the
QGP.

The considered model is here gradually refined to take into account those
physical processes present in the full QCD treatment.
The uniform treatment of all hadronic resonances, without artificial stability
island, is reached by a state-dependent hadron resonance width
$\Gamma_i(T)$ given by the inverse collision time scale in a resonance gas
\cite{Blaschke:2011ry}.

In order to remove this unphysical aspect of the otherwise appealing model
one has to extend the spectral broadening also to the light hadrons and thus
describe their disappearance due to the Mott effect while simultaneously the
quark and gluon degrees of freedom appear at high temperatures due to chiral
symmetry restoration and deconfinement.

In the present contribution we will report results obtained by introducing a
unified treatment of all hadronic resonances with a state-dependent width
$\Gamma_i(T)$ in accordance with the inverse hadronic collision time scale
from a recent model for chemical freeze-out in a resonance gas
\cite{Blaschke:2011ry}.
The appearance of quark and gluon degrees of freedom is introduced by the
Polyakov-loop improved Nambu--Jona-Lasinio (PNJL) model
\cite{Fukushima:2003fw,Ratti:2005jh}.
The model is further refined by adding perturbative corrections to
$\mathcal{O}(\alpha_s)$ for the high-momentum region above the three-momentum
cutoff inherent in the PNJL model.
One obtains eventually a good agreement with lattice QCD data, comparable with
all important physical characteristics taken into account.

\section{Extended Mott-Hagedorn resonance gas}
\subsection{No quarks and gluons; hadronic spectral function with
state-independent ansatz}
We introduce the width $\Gamma$ of a resonance in the statistical
model through the spectral function
\begin{equation}
\label{one}
A(M,m)=N_M \frac{\Gamma \cdot m}{(M^2-m^2)^2+\Gamma^2 \cdot m^2}~,
\end{equation}
a Breit-Wigner distribution of virtual masses with a maximum at $M =
m$ and the normalization factor
\begin{eqnarray}
\label{two}
N_M &=& \left[\, \int\limits_{m_0^2}^\infty {d(M^2)}
\frac{ \Gamma \cdot m }{ ( M^2 - m^2)^2 + \Gamma^2 \cdot m^2  } \right]^{-1}
\nonumber\\
&=&\frac{1}{ \frac{\pi}{2} + \arctan \left( \frac{m^2 - m^2_0 }{ \Gamma
\cdot m} \right) }\,.
\end{eqnarray}
And the model ansatz for the resonance width $\Gamma$ is given by
\cite{Blaschke:2003ut}

\begin{equation}
\label{three}
\Gamma (T) = C_{\Gamma}~ \left( \frac{ m}{T_H} \right)^{N_m}
\left( \frac{ T}{T_H} \right)^{N_T} \exp \left( \frac{ m}{T_H }
\right)~,
\end{equation}
where $C_{\Gamma} = 10^{-4}$ MeV, $N_m = 2.5$, $N_T = 6.5$ and the
Hagedorn temperature $T_H = 165$ MeV.

The energy density of this model with zero resonance proper volume
for given temperature $T$ and chemical potentials: $\mu_B$ for
baryon number and $\mu_{S}$ for strangeness, can be cast in the form
\begin{eqnarray}
\label{four}
\varepsilon(T,\mu_B,\mu_S) &=&
\sum_{i:~ m_i < m_0} g_i ~\varepsilon_i (T,\mu_i;m_i)\nonumber \\
&& \hspace{-3cm}
+ \sum_{i:~ m_i \geq m_0} g_i ~\int\limits_{m_0^2}^\infty {d(M^2)}
~A(M,m_i)~\varepsilon_i (T,\mu_i;M),
%
%\label{one}
\end{eqnarray}
where $m_0 = 1$ GeV and the energy density per degree of freedom
with a mass $M$ is
\begin{equation}
\label{five}
%
%%%\hspace*{2.0cm}
%
\varepsilon_i (T,\mu_i;M)  = \int  \frac{d^3 k}{ (2 \pi)^3 }
\frac{\sqrt{k^2+M^2}}{\exp \left(\frac{\sqrt{k^2 +M^2} - \mu_i}{T}
\right)  + \delta_i } \, ,
\end{equation}
with the degeneracy  $g_i$ and the chemical potential $\mu_i = B_i
\cdot \mu_B + S_i \cdot \mu_S$ of hadron $i$. For mesons,
$\delta_{i} = -1 ~$ and for baryons $~ \delta_{i} = 1$. According to
Eq. (\ref{one}) the energy density of hadrons consists of the
contribution of light hadrons for $m_i < m_0$ ~ and the contribution
of heavier hadrons smeared with the spectral function for $m_i \geq
m_0$.

For simplicity, we assume $n_{S} = 0$ for the strangeness number
density and $n_{B}= 0$ for the baryon number density. Then $\mu_{B}
= 0$ and $\mu_{S} = 0$ always, so the temperature is the only
significant statistical parameter here. In such a case and for a
fixed volume we have
\begin{equation}
\label{six}
\varepsilon + P = T \cdot s = T \cdot \frac{\partial P}{\partial T
}~,
\end{equation}
where $P = P(T)$ and $s = s(T)$ are the pressure and entropy
density, respectively. This is the first order ordinary differential
equation for the pressure and the general solution reads
\begin{equation}
\label{seven}
P(T) = \frac{T}{T_0}\cdot P_0 + T \int\limits_{T_0}^T
{dT'}~\frac{\varepsilon(T')}{T'^2}~,
\end{equation}
where $P_0 = P(T_0)$. To have well-defined solution for the initial
temperature $T_0 = 0$ one has to assume that $\lim_{T_0 \rightarrow
0}P(T_0)/T_0 = s_0 < \infty$. Then
\begin{equation}
\label{eight}
P(T) = s_0 \cdot T + T \int\limits_{0}^T
{dT'}~\frac{\varepsilon(T')}{T'^2}~.
\end{equation}
And the entropy density reads:
\begin{equation}
\label{nine}
s(T) = \frac{\partial P}{\partial T } = s_0 + \int\limits_{0}^T
{dT'}~\frac{\varepsilon(T')}{T'^2} + \frac{\varepsilon(T)}{T}~,
\end{equation}
where $s(0) = s_0$. We put $s_0 = 0$ as suggested by the Nernst
postulate. The sound velocity squared is given by
\begin{equation}
\label{ten}
c_s^2 = \frac{\partial P}{\partial \varepsilon }~.
\end{equation}

\begin{figure}[!th]
\includegraphics[width=0.7\textwidth]{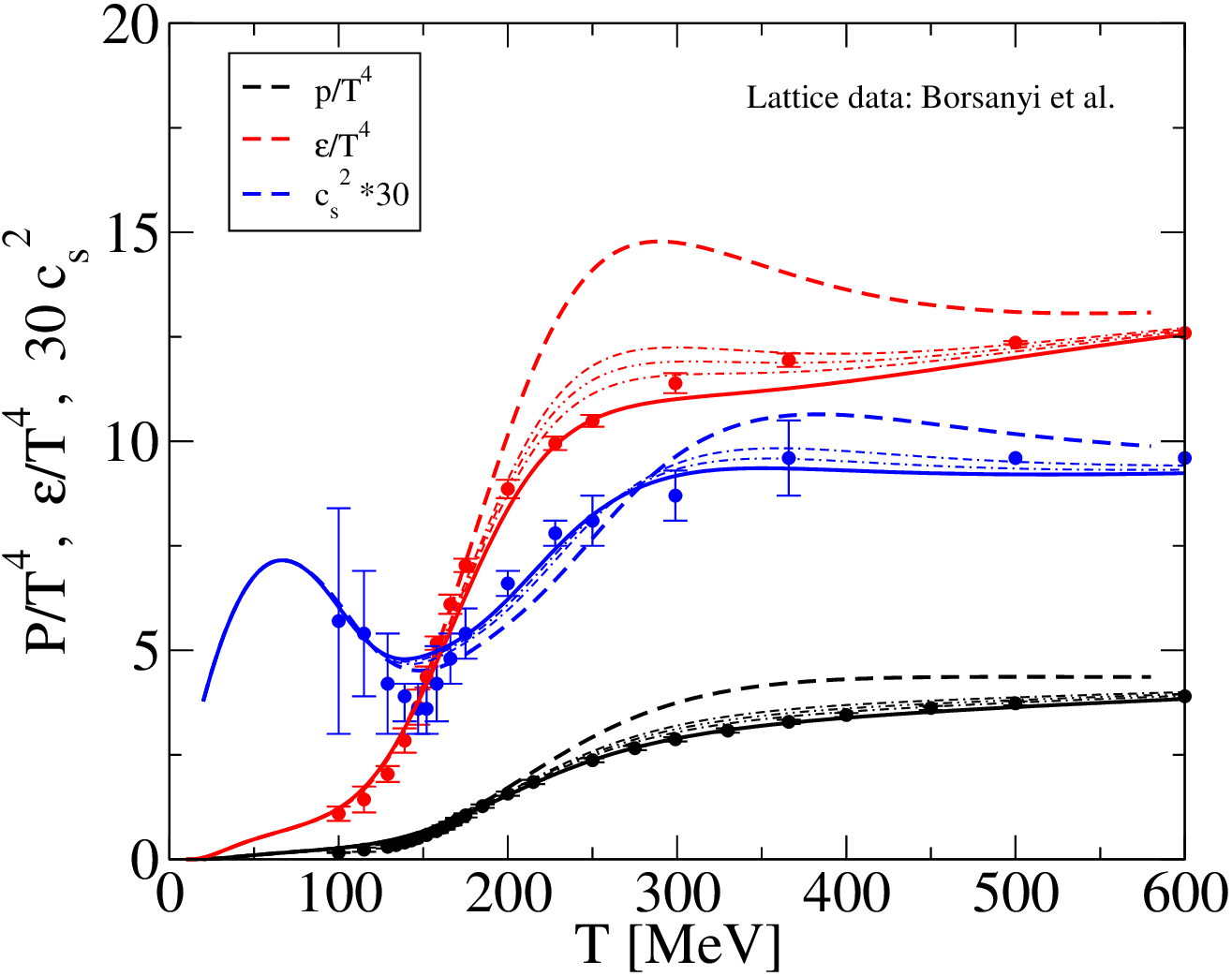}
\caption{\label{Fig.1}
(Color online) Thermodynamic quantities for the old Mott-Hagedorn Resonance Gas model
\cite{Blaschke:2003ut}.
Different line styles correspond to different values for the parameter
$N_m$ in the range from $N_m=2.5$ (dashed line) to $N_m=3.0$ (solid line).
Lattice QCD data are from Ref.~\cite{Borsanyi:2010cj}.
}
\end{figure}

In Fig.~\ref{Fig.1} we show the results for the thermodynamic quantities
(pressure, energy density and squared speed of sound) of the MHRG model
at this stage.
The nice correspondence with results from lattice QCD is not accidental
for the temperature region $T\sim T_c \sim 200$ MeV, where has been shown in
\cite{Borsanyi:2010cj} that a hadron resonance gas perfectly describes the
lattice QCD data.
For $T>T_c$ the broadening of the spectral function (\ref{one}) which affects
at this stage of the model only the hadronic resonances with $m>m_0$ leads
to the vanishing of their contribution at about $2T_c$ while the light
hadrons with masses $m<m_0$ are not affected and gradually reach the Stefan-\-
Boltzmann (SB) limit determined by their number of degrees of freedom.
As has been noted in \cite{Brown:1991dj}, this number
($\sum_{i=\pi,K,\eta,f_0,\rho,\omega,K^*,\eta',f_0,a_0}g_i=
3+4+1+1+9+3+6+1+1+3=32$)
accidentally (or by duality arguments) coincides with that of the quarks and
gluons ($\sum_{i=q,g}g_i=7/8*N_c*N_f*N_s*2 + (N_c^2-1)*2=31.5$) for
$N_c=N_f=3$.
Therefore, imposing that all mesons lighter than $m_0=1$ GeV are stable
provides us with a SB limit at high temperatures which fakes that of quarks
and gluons in the case for three flavors.

Although providing us with an excellent fit of the lattice data, the
high-temperature phase of this model is unphysical since it ignores the
Mott effect for light hadrons. Due to the chiral phase transition at $T_c$,
the quarks loose their mass and therefore the threshold of the continuum
of quark-antiquark scattering states is lowered.
At the same time the light meson masses, however, remain almost unaffected by
the increase in the temperature of the system. Consequently, they merge the
continuum and become unbound - their spectral function changes from a
delta-function (on-shell bound states) to a Breit-Wigner-type (off-shell,
resonant scattering states).
This phenomenon is the hadronic analogue \cite{Zablocki:2010zz} of the
Mott-Anderson transition for electrons in solid state physics
(insulator-conductor transition).

It has been first introduced for the hadronic-to-quark-matter transition in
\cite{Blaschke:1984yj}. Later, within the NJL model, a microscopic approach to
the thermodynamics of the Mott dissociation of mesons in quark matter has been
given in the form of a generalized Beth-Uhlenbeck equation of state
\cite{Hufner:1994ma}, see also \cite{Radzhabov:2010dd}.

\subsection{Hadronic spectral function with state-dependent ansatz}

As a microscopic treatment of the Mott effect for all resonances is presently
out of reach, we introduce an ansatz for a state-dependent hadron resonance
width $\Gamma_i(T)$ given by the inverse collision time scale recently
suggested within an approach to the chemical freeze-out and chiral condensate
in a resonance gas \cite{Blaschke:2011ry}
\begin{equation}
\label{Gamma}
\Gamma_i (T) = \tau_{\rm coll,i}^{-1}(T)
= \sum_{j}\lambda\,\langle r_i^2\rangle_T \langle r_j^2\rangle_T~n_j(T)~,
\end{equation}
which is based on a binary collision approximation and relaxation time ansatz
using for the in-medium hadron-hadron cross sections the geometrical
Povh-H\"ufner law \cite{Povh:1990ad}.
In Eq.~(\ref{Gamma}) the coefficient $\lambda$ is a free parameter, $n_j(T)$ is
the partial density of the hadron $j$ and the mean squared radii of hadrons
$\langle r_i^2 \rangle_T$ obtain in the medium a temperature dependence which
is governed by the (partial) restoration of chiral symmetry.
For the pion this was quantitatively studied  within the
NJL model \cite{Hippe:1995hu} and it was shown that close to the Mott
transition
%the chiral perturbation theory corrections can be safely neglected and
the pion radius is well approximated by
\begin{equation}
r_\pi^2(T)=\frac{3}{4\pi^2} f_\pi^{-2}(T)
=\frac{3M_\pi^2}{4\pi^2m_q}
|\langle \bar{q} q \rangle_{T}|^{-1}~.
\end{equation}
Here the Gell-Mann--Oakes--Renner relation has been used and the pion mass
shall be assumed chirally protected and thus temperature independent.

For the nucleon,  we shall assume the radius to consist of two
components, a medium independent hard core radius $r_0$ and a pion cloud
contribution
%\begin{equation}
$r_N^2(T)=r_0^2+r_\pi^2(T)~,$
%\end{equation}
where from the vacuum values $r_\pi=0.59$ fm and $r_N=0.74$ fm  one gets
$r_0=0.45$ fm.
A key point of our approach is that the temperature dependent hadronic radii
shall diverge when hadron dissociation (Mott effect) sets in, driven basically
by the restoration of chiral symmetry.
%A further simplifying assumption derives from that statement.
As a consequence, in the vicinity of the chiral restoration temperature all
meson radii shall behave like that of the pion and all baryon radii like that
of the nucleon.
%The chiral condensate used in subsequent calculations has a form
%\[-\langle \bar{q} q \rangle_{T} =
%304.8\left[1-\tanh\left(0.002\,T -1\right) \right]\,.\]

\begin{figure}[!th]
\includegraphics[width=0.7\textwidth]{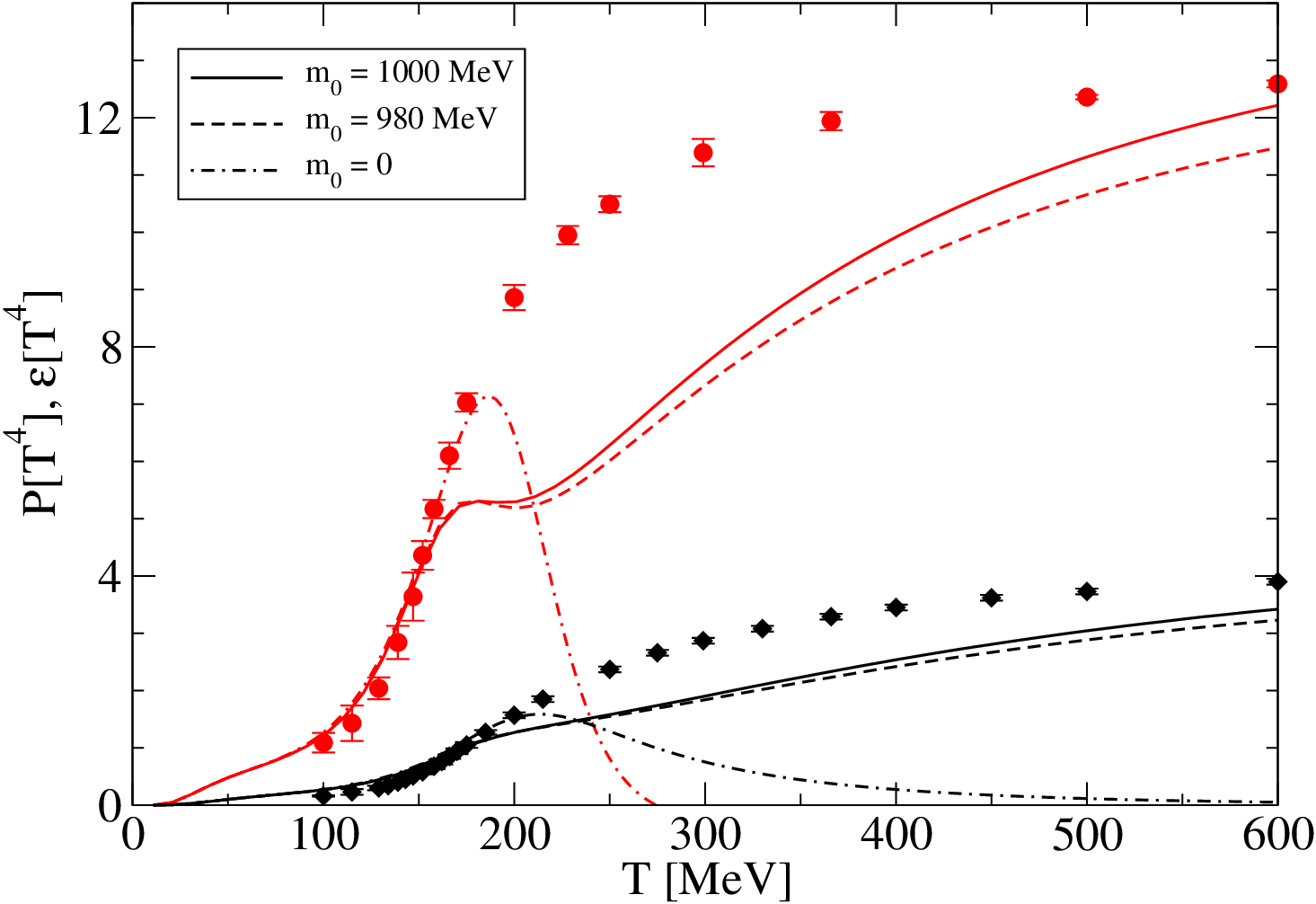}
\caption{\label{Fig.1a} (Color online) Energy density (red lines and symbols)
and pressure (black lines and symbols) for the state-dependent width
model of Eq.~(\ref{Gamma}) and three values of the mass threshold $m_0$:
1 GeV (solid lines), 980 MeV (dashed lines), 0 (dash-dotted lines).
Lattice QCD data are from Ref.~\cite{Borsanyi:2010cj}.}
\end{figure}

The resulting energy density behaviour is shown in Fig.~\ref{Fig.1a}.
This part of the model we call Mott-Hagedorn-Resonance-Gas (MHRG).
When all hadrons are gone at $T\sim 250$ MeV, we are clearly missing degrees
of freedom!

\section{Quarks, gluons and hadron resonances}

We improve the PNJL model over its standard versions
\cite{Fukushima:2003fw,Ratti:2005jh} by adding perturbative corrections in
$\mathcal{O}(\alpha_s)$ for the high-momentum region above the three-momentum
cutoff $\Lambda$.
In the second step, the MHRG part is replaced by its final form, using the
state-dependent spectral function for the description of the Mott dissociation
of all hadron resonances above the chiral transition.
The total pressure obtains the form
\begin{equation}
P(T)=P_{\rm MHRG}(T)+P_{\rm PNJL,MF}(T)+P_2(T) ~.
\end{equation}
where $P_{\rm MHRG}(T)$ stands for the pressure of the MHRG model, accounting
for the dissociation of hadrons in hot dense, matter.

The $\mathcal{O}(\alpha_s)$ corrections can be split in quark and gluon
contributions
\begin{equation}
\label{P2}
P_2(T)=P_2^{{\rm quark}}(T) + P_2^{{\rm gluon}}(T)~,
\end{equation}
where $P_2^{{\rm quark}}$  stands for the quark contribution and
$P_2^{{\rm gluon}}$ contains the ghost and gluon contributions.
The total perturbative QCD correction to $\mathcal{O}(\alpha_s)$ is
\begin{equation}
P_2=-\frac{8}{\pi}\alpha_s T^4(I_{\Lambda}^+
+\frac{3}{\pi^2}((I_{\Lambda}^+)^2+(I_{\Lambda}^-)^2)),
\end{equation}
where
$I^{\pm}_{\Lambda}=\int\limits_{\Lambda/T}^{\infty}\frac{{\rm d}x~x}{{\rm e}^x\pm 1}$.
The corresponding contribution to the energy density is given in standard way
by Eq.~(\ref{six}).

We will now include an effective description of the dissociation of hadrons
due to the Mott effect into the hadron resonance gas model by including the
state dependent hadron resonance width (\ref{Gamma}) into the definition of the
HRG pressure
\begin{equation}
P_{\rm MHRG}(T)=\sum_{i}\delta_id_i\!\int\!\frac{d^3p}{(2\pi)^3}dM\,
A_i(M) T \ln\left(1+\delta_i{\rm e}^{-\sqrt{p^2+M^2}/T} \right)\,.
\end{equation}
From the pressure as a thermodynamic potential all relevant thermodynamical
functions can be obtained.
Combining the $\alpha_s$ corrected meanfield PNJL model for the quark-gluon
subsystem with the MHRG description of the hadronic resonances we obtain the
results shown in the right panel of Fig.~\ref{Fig.1} where the resulting
partial contributions in comparison with lattice QCD data from
Ref.~\cite{Borsanyi:2010cj} are shown.

We see that the lattice QCD thermodynamics is in full accordance with a
hadron resonance gas up to a temperature of $\sim 170$ MeV which corresponds
to the pseudocritical temperature of the chiral phase transition.
The lattice data saturate below the Stefan-Boltzmann limit of an ideal
quark-gluon gas at high temperatures.
The PNJL model, however, attains this limit by construction.
The deviation is to good accuracy described by perturbative corrections to
$\mathcal{O}(\alpha_s)$ which vanish at low temperatures due to an infrared
cutoff procedure.
The transition region $170\le T[{\rm MeV}]\le 250$ is described by the MHRG
model, resulting in a decreasing HRG pressure which vanishes at $T \sim 250$
MeV.

\begin{figure}[!th]
\includegraphics[width=0.7\textwidth]{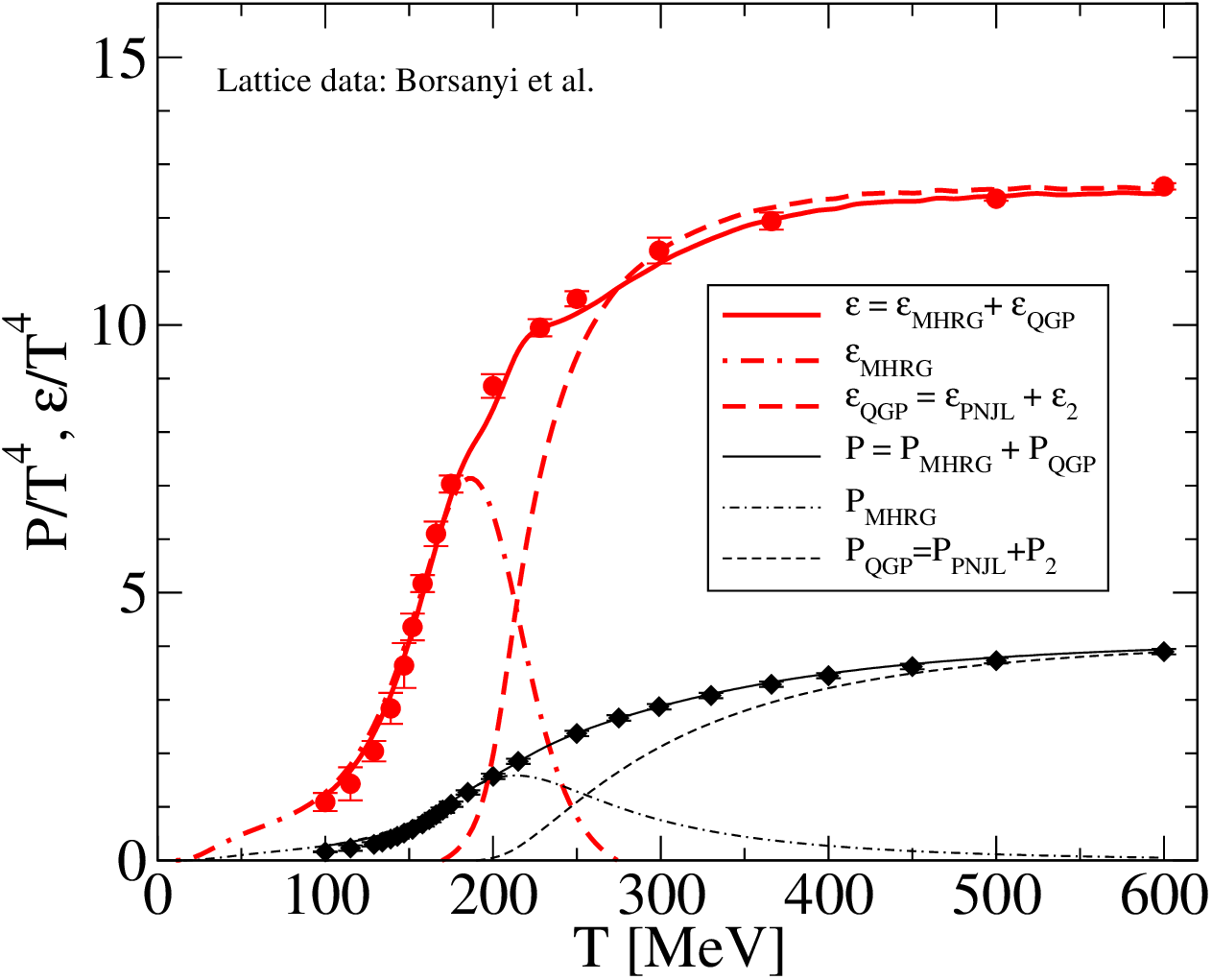}
\caption{\label{Fig.3}
(Color online) Thermodynamic quantities for the new Mott-Hagedorn Resonance Gas where
quark-gluon plasma contributions are described within the PNJL model
including $\alpha_s$ corrections (dashed lines).
Hadronic resonances are described within the resonance gas with finite width,
as an implementation of the Mott effect (dash-dotted line).
The sum of both contributions (solid lines) is shown for the energy density
(thick lines) and pressure (thin lines) in
comparison with the lattice data from \cite{Borsanyi:2010cj}.
}
\end{figure}

We have presented two stages of an effective model description of QCD
thermodynamics at finite temperatures which properly accounts for the fact
that in the QCD transition region it is dominated by a tower of hadronic
resonances.
To this end we have further developed a generalization of the Hagedorn
resonance gas thermodynamics which includes the finite lifetime of hadronic
resonances in a hot and dense medium by a model ansatz for a temperature- and
mass dependent spectral function.

\section{Conclusion and outlook}
\begin{figure}[h!]
\includegraphics[width=0.7\textwidth]{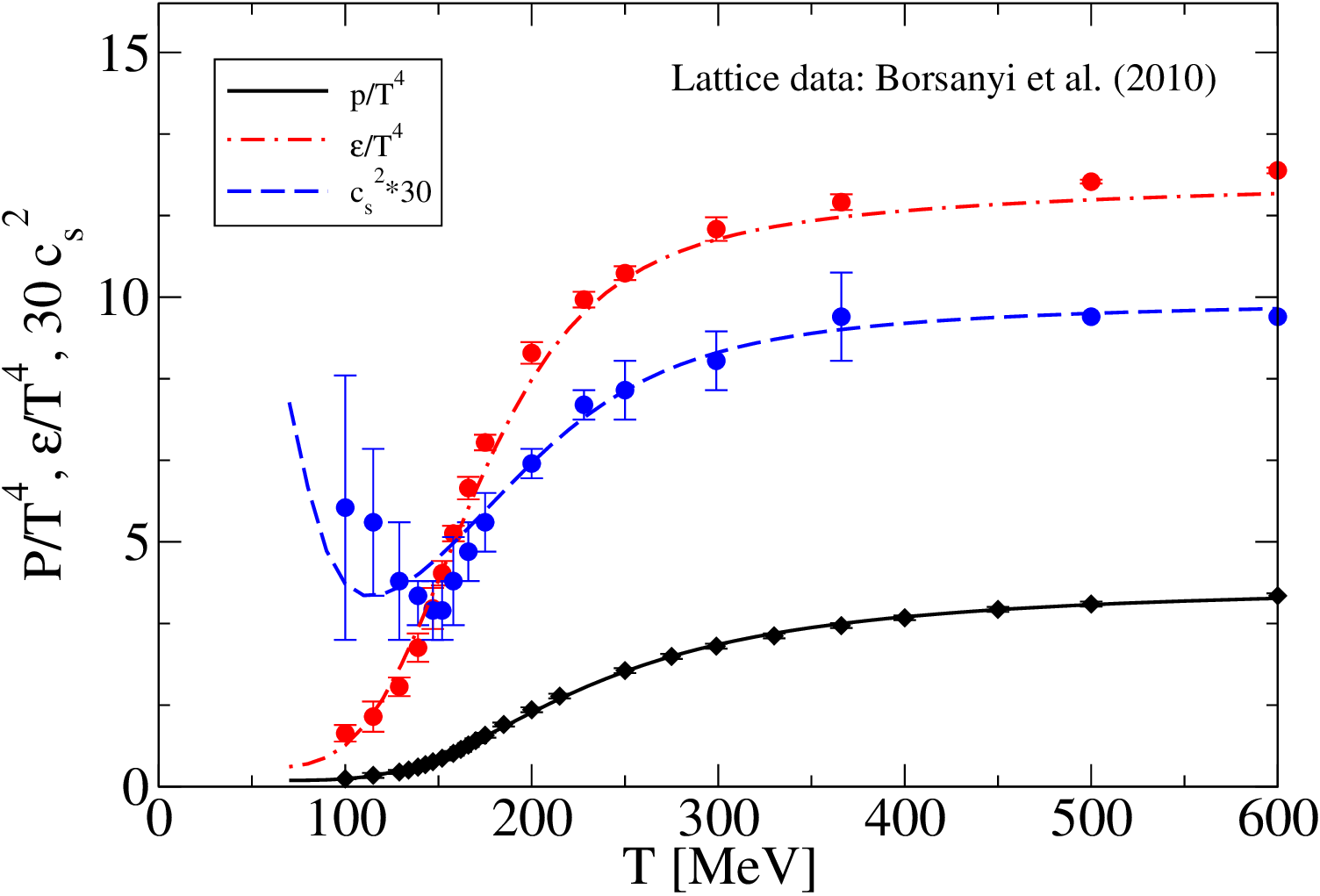}
\caption{\label{Fig.3}(Color online) Thermodynamic quantities as in Fig.~\ref{Fig.1} for the
MHRG-PNJL model compared to lattice data from Ref.~\cite{Borsanyi:2010cj}.
}
\end{figure}

After having presented the MHRG-PNJL model with the state-dependent spectral
function approach we show the summary of the thermodynamic quantities in
 Fig.~\ref{Fig.3}.
We have presented two stages of an effective model description of QCD
thermodynamics at finite temperatures which properly accounts for the
fact that in the QCD transition region it is dominated by a tower of hadronic
resonances.
In the first of the two stages of the developments we presented here, we have
used the fact that the number of low-lying mesonic degrees of freedom with
masses below $\sim 1$ GeV approximately equals that of the thermodynamic
degrees of freedom of a gas of quark and gluons.
In the second one we have further developed a generalization of the Hagedorn
resonance gas thermodynamics which includes the finite lifetime of heavy
resonances in a hot and dense medium by a model ansatz for a temperature- and
mass dependent spectral function which is motivated by a model for the
collision time successfully applied in the kinetic description of chemical
freeze-out from a hadron resonance gas.
A next step should take into account also the effects of continuum correlations
in hadronic scattering channels in accordance with the Levinson theorem
\cite{Dashen:1969ep}
as discussed recently for the example of pion dissociation within the
PNJL model \cite{Wergieluk:2012gd}.
Our account for $\mathcal{O}(\alpha_s)$ corrections from quark and gluon
scattering in the plasma may be seen from this perspective.

\ack
D.B. wants to thank K.A. Bugaev, K. Redlich and A. Wergieluk for discussions
and collaboration.
This work was supported in part by the Polish National Science Center (NCN)
under contract No. N N202 0523 40 and the ``Maestro'' programme  DEC-2011/02/A/ST2/00306 as well as by the Russian Foundation for Basic Research under grant No. 11-02-01538-a
(D.B.).
\\[5mm]

\end{document}